\documentclass[a4paper,12pt]{article}

\usepackage[utf8]{inputenc}
\usepackage{amsmath}
\usepackage[hidelinks]{hyperref}
\usepackage{cite}
\usepackage{german}
\usepackage{color}
\usepackage{nicefrac}
\usepackage{graphicx}
\usepackage{subcaption}
\usepackage{listings}
\usepackage{xcolor}
\definecolor{code_color}{RGB}{230,230,230}

\title{A simple simulation of quantum like behavior with a classical oscillator / Einfache Simulation von Quantensystemen mittels eines klassischen Oszillator}

\author{M.Bühler}

\begin{document}

\begin{center}
\Large 		Einfache Simulation von Quantensystemen 		\\
		mittels eines klassischen Oszillator 			\par \vspace{3mm}
\footnotesize 	M.Bühler, buehler.xy@t-online.de 			\par
		6. Dezember 2017
\end{center}



\renewcommand\abstractname{Zusammenfassung} 
\begin{abstract}
 
Seit Jahren ist ein klassischer Oszillator bekannt\footnote{bedauerlicherweise ist es dem Autor nicht gelungen eine wissenschaftliche Referenz zu diesem Oszillatortyp zu finden}, der es erlaubt einige grundlegende Eigenschaften von quantenmechanischen Systemen darzustellen, ohne die anspruchsvolle Mathematik der Quantenmechanik benutzen zu müssen. Dadurch ist es möglich in der Schule und Hochschule in Kursen zur Einführung in die Quantenmechanik eine intuitive Vorstellung zur Quantenmechanik zu entwickeln und Eigenschafen wie spontane und stimulierte Emission, Quantisierung und Kollaps des Oszillator analog zum Kollaps der Wellenfunktion sowie die zwanglose Einführung der Wahrscheinlichkeitsinterpretation anschaulich darzustellen. Die Simulation des Oszillator mittels der GNU Octave Software und dem zugehörigen Quelltext wird gezeigt.
\end{abstract}
%
%
\section{\large Oszillatormodell}
Der klassische harmonische Oszillator mit konstanter Dämpfung ist seit langem bekannt\cite{demtroeder2015expphys}-\nocite{steinhauser2017qmfnatwiss} \cite{pade2012qmzufuss} und wird gerne als klassisches Analog zur Quantentheorie verwendet um Ähnlichkeiten und Unterschiede zu beleuchten. Für Bemühungen die Didaktik zur Einführung der Quantenmechanik zu verbessern \cite{kueblbeck2002wesenszuege} sind allerdings Beispiele mit nur limitierter Ähnlichkeit nicht uneingeschränkt hilfreich, weil der ``Sprung'' zur Quantenmechanik sehr groß bleibt. Wenig bekannt ist hingegen, dass derselbe Oszillator mit einer nur unwesentlichen Verallgemeinerung zur Darstellung viel weiterreichender quantenmechanischer Eigenschaften geeignet ist, obwohl es sich weiterhin um ein völlig klassisches System handelt. Dabei bleibt der Oszillator leicht verständlich und erlaubt ein intuitives Verständnis seiner Eigenschaften. Der entscheidende Punkt ist, dass die Dämpfung $D$ so geändert wird, dass eine ``Quantisierung'' der Oszillation künstlich erzeugt wird. Dies kann in sehr allgemeiner Weise erfolgen, indem der Dämpfungsterm $D(E)$ als Funktion der Energie $E$ des Oszillator betrachtet wird, wobei neben der Stetigkeit, wie üblich, noch $D(E)\geq 0$ gefordert wird. Dadurch ergibt sich folgende Oszillator Gleichung ($m$: Masse; $f$: Federkonstante), mit einer zeitlichen Anregung $a(t)$,
\begin{equation}
\label{eq:Oszi}
 m \ddot{x}+ D(E)\dot{x}+ f x =a(t);\qquad E: \mathrm{Energie\ d.\ Oszillator}
\end{equation}
welche offensichtlich den Oszillator mit konstanter Dämpfung als Spezialfall enthält. Interessant sind jetzt aber gerade jene Fälle mit einem Dämpfungsterm der eine oder mehrere Nullstellen hat, welche genau die stabilen Energieniveaus $E_n$ des Oszillator beschreiben.
\begin{equation}
\mathrm{stabile\ Energieniveaus :}\qquad E_n= \{E\  |D(E)=0 \}
\end{equation}
Dies ist leicht einzusehen indem man zuerst einen Oszillator mit beliebiger Energie aber ohne Anregung $a(t)=0$ betrachtet. Da die Dämpfung im allgemeinen $D>0$ ist, wird der Oszillator so lange seine Amplitude verringern, bis er sich dem nächst niederen $E_n$ immer mehr annähert, ohne dieses Niveau wegen der abnehmenden Dämpfung zu erreichen. Dieser Zustand hat folglich eine unendliche Lebensdauer $\tau$. Wird in diesem Zustand nun ein beliebig kleines Rauschen hinzugefügt, also $a(t)\neq0$, so wird der Zustand nicht stabil bleiben, sondern mit gewisser Wahrscheinlichkeit das erreichte Energieniveau unterschreiten, wodurch die Dämpfung wieder einsetzt und der Zustand weiter zerfällt.
\begin{figure}[h]
\centering
  \begin{subfigure}[b]{0.48\textwidth}
        \includegraphics[width=\textwidth]{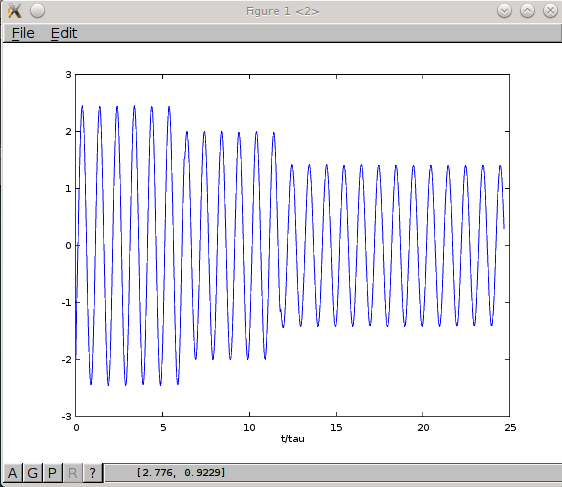}
        \caption{Kollaps des Oszillator}
        \label{Oszi}
  \end{subfigure}
    ~ 
  \begin{subfigure}[b]{0.48\textwidth}
        \includegraphics[width=\textwidth]{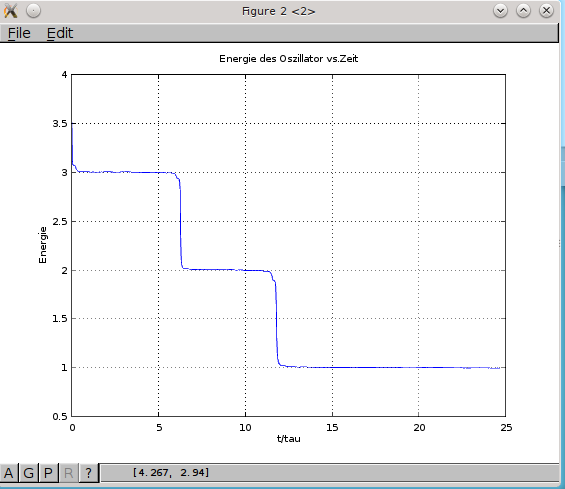}
        \caption{``quantisierte'' Energie}
        \label{Energie Oszi}
  \end{subfigure}  
  \caption{Simulation des Oszillator mit Rauschen. a): Kollaps des Oszillator erfolgt nicht vorhersehbar. b) Energie des Oszillator aus a) ändert sich ``quantisiert''.}
  \label{fig:Simulation}
\end{figure}
Das hinzugefügte Rauschen verursacht also eine endliche Lebensdauer $\tau$ der Energieniveaus und entspricht der spontanen Emission in der Quantenmechanik, wo ebenfalls das unvermeidliche Rauschen aufgrund der Heisenberg'schen Unschärfe zur endlichen Lebensdauer der Zustände führt\cite{schwabl1990quant}. Damit ist aber auch verständlich, dass jede zusätzliche Anregung unabhängig davon ob es sich um Rauschen oder eine absichtliche Anregung handelt die Lebensdauer eines Energieniveaus verkürzt, wobei natürlich Anregungen, welche nahe bei der Resonanzfrequenz $\omega_0 =\sqrt{^f/_m}$ des Oszillator liegen, effektiver sind. Daher wird in diesem Oszillatormodell stimulierte Emission bzw. der Zerfall eines Energieniveaus durch äußere Anregung auf die gleiche Weise wie in der Quantenmechanik erklärt. Wenn der Übergang von einem Energieniveau $E_n$ zum nächst niederen Niveau durch einen Rauschbeitrag bestimmt wird, ist dieser Zeitpunkt wegen der stochastischen Natur des Rauschens nicht berechenbar, solange das Rauschen nicht exakt bekannt ist, und daher ist der Übergang von einem Niveau zum nächsten nur als Wahrscheinlichkeitsdichte realisierbar und wird neben dem Rauschen vom Dämpfungsterm bestimmt. Ist die Lebensdauer eines Energieniveau länger als der Übergang, so kollabiert das Energieniveau tatsächlich zu einem unvorhersehbaren Zeitpunkt und veranschaulicht so den Kollaps einer Wellenfunktion (Abb.\ref{fig:Simulation}), genau so wie in der gängigen Interpretation der Quantenmechanik.
\par Da der Dämpfungsterm die Oszillatorgleichung \eqref{eq:Oszi} zu einer nichtlinearen partiellen Differentialgleichung zweiter Ordnung macht, ist eine Simulation des Oszillator, wie im folgenden Kapitel gezeigt, die einfachste Methode um seine Eigenschaften kennenzulernen.
%
%
\section{\large Simulation der Oszillatorgleichung}
Die Simulation der Oszillatorgleichung kann mit Hilfe einer geeigneten Software leicht durchgeführt werden und ist im folgenden beispielhaft mit der GNU Octave Software gezeigt. Wie üblich wird dazu die Differentialgleichung zweiter Ordnung in zwei Differentialgleichengen erster Ordnung mit den neuen Variablen $y_1=x$ und $y_2=\dot y_1$ umgeformt. Damit wird aus Gleichung \eqref{eq:Oszi} das folgende Gleichungssystem:
\begin{equation}
  \begin{array}{r@{\ }l}
    \dot{y}_1   &= y_2  \\[4mm]
    \dot{y}_2	&= -\dfrac{f}{m} y_1 - \dfrac{D(E)}{m} y_2  + \dfrac{a(t)}{m}
  \end{array}
\end{equation}
Um die gewünschte ``Quantisierung'' der Lösungen zu erreichen wurde als Dämpfungsterm beispielhaft $D(E)=Sin(\pi E)^2$ verwendet, welcher Nullstellen für $E_n=0,1,\dots$ besitzt und damit ein Energiespektrum vergleichbar mit dem harmonischen Oszillator in der Quantenmechanik. Die Dämpfungsfunktion kann dabei sehr frei gewählt werden (natürlich unter der Bedingung $D(E)\geq 0$), da ganz wesentlich nur die Nullstellen für die Lage der stabilen Niveaus verantwortlich sind. Die Energie $E$ des Oszillator wird aus der kinetischen und potentiellen Energie bestimmt. Im unten gegebenen Quelltext wurde als Anregung ein normalverteiltes Rauschen verwendet. Zu beachten ist, dass das Rauschen nicht mit dem Funktionsaufruf der Differentialgleichung erzeugt werden kann, da sonst wegen der adaptiven Anpassung der Schrittweite die Wahrscheinlichkeitsverteilung verändert wird.
\begin{figure}[h]
  \centering
  \begin{subfigure}[b]{0.47\textwidth}
        \includegraphics[width=\textwidth]{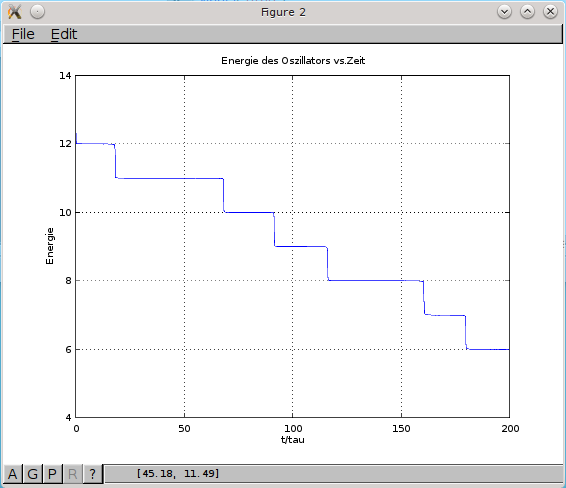}
        \caption{$\sigma=1.0e^{-3}$}
        \label{fig:gull}
  \end{subfigure}
    ~ 
  \begin{subfigure}[b]{0.47\textwidth}
        \includegraphics[width=\textwidth]{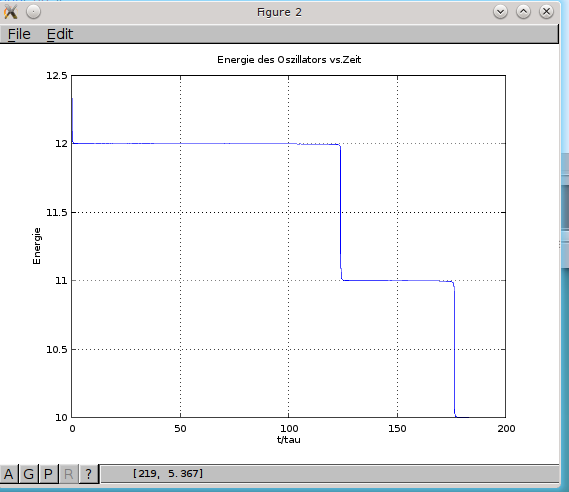}
        \caption{$\sigma=1.0e^{-4}$}
  \end{subfigure}  
  \caption{``Stimulierte Emission'': Änderung des Rauschens mit einer Standard Normalverteilung als Anregung ändert die Lebensdauer der Zustände. (a) und (b) haben gleiche Zeitachse!}
  \label{sigmatest}
\end{figure}
Daher wurde das Rauschen für die geforderte Zeitspanne vorher erzeugt und  durch die Wahl des minimalen $\Delta t$ die maximale Frequenz des Rauschens vorgegeben. Dieses diskrete Rauschen ergibt dann mittels Interpolation die Anregungsfunktion $a(t)$. Weiter ist zu beachten, dass die relative und absolute Genauigkeit so gewählt werden, dass Energiezustände mit sehr langer Lebensdauer nicht aufgrund der endlichen Genauigkeit bereits beschränkt werden, da die Genauigkeit nichts anderes als einen entsprechenden Rauschbeitrag darstellt. Dies kann einfach getestet werden indem die Differentialgleichung ohne den Anregungsterm gelöst und dann die Genauigkeit solange erhöht wird, bis die gewünschte Lebensdauer erreicht ist. Natürlich hat dies den Nachteil, dass sich die Rechenzeit drastisch erhöht, zumal wenn dann zusätzlich ein Rauschbeitrag verwendet wird.
\pagebreak


\lstset{language=Octave, backgroundcolor=\color{code_color}, basicstyle=\footnotesize, stringstyle=\color{red}, keywordstyle=\bfseries\color{blue}}

\lstset{ morekeywords={lsode_options,normrnd}}

\begin{lstlisting}
% Oszillatorgleichung  mit lsode loesen
% M d^2y1/dt^2 -D(E) dy1/dt -F y1 = Anregung(t) 
% Umformung zu 2 DGL 1. Ordnung: y2=dy1/dt  
%      dy1/dt=      0 y1(t)	+        y2(t)
%      dy2/dt=    F/M y1(t)	+ D(E)/M y2(t)   +  1/M an(t)

function Y=Oszillatorsim()
    global 	M F KOP SIGMA FMAX TMAX Tvek Rvek
    F=1;		%Federkonstante
    M=1;		%Masse
    KOP=0.5;		%Kopplung  der Daempfung
    SIGMA=1.5e-3	%Sigma fuer normalverteiltes Rauschen
    FMAX=20;		%maximale Frequenz des Rauschens
    TMAX=150;		%maximale Zeit
%    
% Rauschen,ti und a(ti) initialisieren 	
%
	Tvek=0:1/FMAX:TMAX+3/FMAX; 	% +3ti wg. Interpolation
	Rvek=normrnd(0,SIGMA/sqrt(FMAX),size(Tvek));
	t_interval=[0:0.2:TMAX];	%Punkte zum Plotten
%	
% Anfangsbedingung Phase y und dy/dt zufaellig, Energie E0
%
	E0=5.1;			%Start Energie des Oszi.
	ph=rand()*2*pi;	
	anf=[sin(ph),cos(ph)]'*sqrt(2*E0);
%	
% Genauigkeit der Simulation setzen, DGL berechnen	
%
	lsode_options("absolute tolerance",1.0e-8);
	lsode_options("relative tolerance",1.0e-8);
	Y=lsode("sys",anf,t_interval);
	
% Ergebnis darstellen	
	
	tau=(2*pi)/sqrt(F/M);		%skalieren mit Periode
	figure(1);
	plot(t_interval/tau,Y(:,2));
	xlabel("t/tau");
	figure(2);
	E=((Y.^2.)*[F,M]')*0.5;		% Energie des Oszillator
	plot(t_interval/tau,E);
	title("Energie des Oszillator vs.Zeit");
	grid("on");
	xlabel("t/tau");
	ylabel("Energie");
	drawnow();
	
endfunction
\end{lstlisting}
\pagebreak
Die verwendeten Funktionen für die Differentialgleichung, Dämpfung und Anregung sind dabei:
\begin{lstlisting}
% DGL System
%
function dydt=sys(y,t)
	global F M 
	dydt=zeros(2,1);
	dydt(1)=y(2);
	dydt(2)=(-F*y(1)-R(y)*y(2)+ an(t))/M;
endfunction
%
% Daempfung
%
function r=R(y)
	global F M KOP
	E=[F,M]*(y.^2.)*0.5;	% Energie des Oszillator
	r=KOP*sin(pi*E)^2;	% D(E)~sin[pi E)^2; En=0,1,2...
endfunction
%
% Anregung
%
% interpoliert Datenpunkte um Rauschen zu erzeugen
% Datenpunkte ti in Tvek, yi in Rvek;  dt konstant
% fuer beliebige Anregung kann hier f(t) addiert werden

function y=an(t)
% t Zeitpunkt
	global Tvek Rvek
	
	y=interp1(Tvek,Rvek,t,"*spline");	% +f(t);
endfunction
	
\end{lstlisting}
%
%
\section{\large Anmerkungen zur Physik}
Gelegentlich wird der Eindruck erweckt, dass quantenmechanische Systeme klassisch nicht erklärbar sind. Dazu zählen auch spontane und stimulierte Emission, der Kollaps der Wellenfunktion sowie die Tatsache, dass nur Wahrscheinlichkeiten für die Messwerte angegeben werden können. Derartige Behauptungen werden aber meist ohne formalen Beweis vorgebracht und sind, wie das Beispiel dieses Oszillatormodells zeigt, nicht immer richtig. Dies führt dazu, dass die Trennlinie zwischen klassischer und Quantenphysik nicht scharf definiert ist, und insoweit macht es Sinn diese Trennlinie bisweilen einer genauen Untersuchung zu unterziehen.

Das hier vorgestellte Oszillatormodell verschiebt den klassischen Bereich in Richtung Quantenphysik, indem sie durch die Quantisierung der Energieniveaus von Oszillatoren etwa den Vorstellungen von Max Planck\cite{planck1900normalspek} entspricht. Dabei wird angenommen, dass Atome durch quantisierte Oszillatoren repräsentiert werden, nicht aber das zugehörige elektromagnetische Feld, welches für die abgegebene oder aufgenommene Energie verantwortlich ist. Im Oszillatormodell wird die Quantisierung dynamisch erreicht und bei angenommener Energieerhaltung werden beliebig kleine Energien auf das elektromagnetische Feld übertragen. Gleichzeitig ist aber die Quantisierung dafür verantwortlich, dass ein nicht zu verhindernder  Rauschbeitrag existiert. Dazu muss man sich nur den Oszillator in einem geschlossenen Hohlraum mit vollständig reflektierenden Wänden vorstellen und annehmen, dass die im Feld vorhandene Energie nicht exakt den Energieniveaus entspricht. Dann folgt zwangsläufig, dass immer ein nicht absorbierbarer Feldbeitrag existiert. Wird dieser Feldbeitrag als zufällig angenommen so entspricht er im Mittel dem der Quantentheorie von $\nicefrac{1}{2} E_n$. Im Gegensatz zur Quantenfeldtheorie \cite{schwabl1990quant} werden hier Phänomene wie die spontane Emission ohne die Quantisierung des elektromagnetischen Feldes möglich und Nichtquantisierung ist im klassischen Fall sogar die Voraussetzung um den zufälligen Zerfall des Oszillator zu erreichen. Der klassische Oszillator erfüllt aber nicht notwendig die Frequenz-Energie Beziehung $E=h\nu$ der Quantentheorie und diese Beziehung muss als Eigenschaft der Oszillatoren formuliert werden, wenn sie gelten soll. Damit widerspricht das klassische Modell der Quantenfeldtheorie klar in Bezug auf die Quantisierung des elektromagnetischen Feldes. Effektiv betrifft dieser Widerspruch aber eher die Interpretation der Quantenfeldtheorie, da sowohl in der Quantenfeldtheorie als auch im hier vorgestellten klassischen Oszillator Emission und Absorption von Energie nur quantisiert erfolgen und damit der Wirkung von Erzeugungs- und Vernichtungsoperatoren entsprechen. Dass die Erzeugungs- und Vernichtungsoperatoren in der Quantenfeldtheorie durch Quantisierung des elektromagnetischen Feldes hergeleitet werden, ändert nichts an der Tatsache, dass der Formalismus der Quantenfeldtheorie zwingend nur die quantisierte Emission und Absorption beschreibt. Wie das Oszillatormodell zeigt, kann die Quantisierung von Emission und Absorption auch erreicht werden, indem nur der Oszillator quantisiert wird und liefert dann zumindest qualitativ gleiche Ergebnisse. Der einzige Bereich wo die Quantenfeldtheorie Aussagen über das elektromagnetische Feld ohne Emission oder Absorption macht betrifft die Vakuumenergie, dort aber ist die Übereinstimmung mit dem Experiment  nicht sehr zufriedenstellend\cite{weinberg1989cosmological}. Weitere Schnittpunkte betreffen die Diskussion zur Interpretation des Kollaps der Wellenfunktion\cite{neumann1996grundlagen}, die im klassischen Modell auf triviale Weise beantwortet wird, die Interpretation von Doppelspalt Experimenten, den Welle-Teilchen Dualismus, die Wahrscheinlichkeitsinterpretation oder den Quanten-Zeno Effekt, der im Modell einfach nachvollzogen werden kann.

\bibliographystyle{plain}   	  

%
%

\end{document}